\documentclass[12pt,tightenlines,%twocolumn,
onecolumn, preprintnumbers, amsmath, amssymb,  prd,  showpacs,
showkeys ]{revtex4}
\usepackage{graphicx}%
\usepackage{amsmath}
\usepackage[normalem,normalbf]{ulem}
\usepackage{amssymb}
\usepackage{bm}
\usepackage{enumerate}
\newcommand{\nn}{\nonumber}
\begin{document}
%-------------------------%-------------------------------------
\title{Blackbody radiation in $\kappa-$Minkowski spacetime}
%------------------------------------------------------------
\author{Hyeong-Chan Kim$^1$}
\email{hckim@phya.yonsei.ac.kr}
%------------------------------------------------------------
\author{Chaiho Rim$^2$}
\email{rim@chonbuk.ac.kr}
%-------------------------------------------------
\author{Jae Hyung Yee$^1$}%
\email{jhyee@yonsei.ac.kr}
%-------------------------------------------------------------
\affiliation{$^1$ Department of Physics, Yonsei University,
Seoul 120-749, Republic of Korea
\\
and
\\
%}%
%-------------------------------------------------------------
%\affiliation{
$^2$ Department of Physics and Research Institute
of Physics and Chemistry, Chonbuk National University,
Jeonju 561-756, Korea.
}%
%------------------------------------------------------------
%\date{\today}%
%-------------------------------------------------------
\bigskip
%-------------------------------------------------------------
\begin{abstract}
%----------------------------------------------------------
\bigskip
We have computed the black body radiation spectra in $\kappa-$Minkowski
space-time, using the quantum mechanical picture of massless scalar
particles as well as effective quantum field theory picture. The black
body radiation depends on how the field theory (and thus how the
$\kappa-$Poincar\'e algebra) handles the ordering effect of the
noncommutative space-time. In addition, there exists a natural momentum
cut-off of the order $\kappa$, beyond which a new real mode takes its
shape from a complex mode and the old real mode flows out to be a new
complex mode. However, the new high momentum real mode should  not be
physical since its contributions to the black-body radiation spoils the
commutative limit.

%
%----------------------------------------------------------
\end{abstract}
%------------------------------------------------------
%--------------------------------------------------------
%------------------------------------------------------------
\maketitle
%---------------------------------------------------------

%%%%%%%%%%%%%%%%%%%%%%%%%%%%%%%%%%%%%
\section{Introduction}
%%%%%%%%%%%%%%%%%%%%%%%%%%%%%%%%%%%%%
In the last decade, there have been attempts to explain the cosmic
observational data~\cite{good} as a quantum gravitational effect, and
several non-commutative theories based on the uncertainty of Planck
scale position~\cite{doplicher}, on the twist
formulation~\cite{wess,chaichian} and on the Lie algebraic
deformation~\cite{okon}, have been proposed as a theoretical framework
to study such effects.
The theories, however, show some unsatisfactory points, such as a
strong fine-tuning problem  at one-loop level~\cite{finetuning}
in Moyal deformed theory
when applied to the high energy scattering theory,
and complex metric problem when applied to gravity theory
directly~\cite{chamseddine}.
When applied to space-time noncommutative field theories,
unitarity of the theory is in question
\cite{unitarity1,unitarity2}. This suggests that one needs to
understand the structure of the quantum field theories in
noncommutative space-times more clearly, find the limits and
differences of the results from the usual commutative field theory
approaches, and investigate the possibility of
explaining the observational data.%

One of the candidates to study the quantum gravity effect on the cosmic
observational results is the $\kappa-$Minkowski space-time~\cite{kappa}
since this deformation respects rotational symmetry. The space-time
coordinates do not commute with each other and satisfy the commutation
relations,
\begin{equation} \label{k-Minkowski}
~[\hat x^0, \hat x^i] = \frac{i}{\kappa} \hat x^i\,,\qquad ~[\hat x^i,
\hat  x^j]=0\,,\qquad i,j=1,2,3\,
\end{equation}
where $\kappa$ is a positive parameter which represents the
deformation of the space-time, whose natural choice is to put
$\kappa=M_P$, the Planck mass \cite{kosinski,glikman,daszkie}.
This $\kappa-$Minkowski space-time gives the dual picture in terms of
the $\kappa-$deformed Poincar\'{e} algebra~\cite{majid}.
When the dimensional parameter $1/\kappa$ is  suppressed, the deformed
Poincar\'{e} symmetry reduces to the commutative limit.

One may construct its differential calculus
where arises a fifth dimension naturally as shown in~\cite{sitarz,gonera}.
In momentum space this turns out to give
the  $4-$dimensional de Sitter space:
\begin{eqnarray*}
({\cal P}_0)^2-({\cal P}_i)^2 -({\cal P}_5)^2 = -\kappa^2,
\end{eqnarray*}
where ${\cal P}_A$ denotes appropriately normalized momentum in
$5-$dimensional momentum space. (See Eq.~(\ref{inv_chimu}) below).
Equipped with the differential structure,
scalar field theory has been studied in $\kappa-$Minkowski
space-time~\cite{kosinski,freidel2,kim0}.
The $\kappa-$deformation is extended to the curved space with
$\kappa-$Robertson-Walker metric and is applied to cosmic microwave
background radiation in~\cite{kim2}.

The black-body radiation effect is another physical example to test the
theory. The case with Moyal deformed space has appeared in
~\cite{fatollahi} and the deformed effect is shown to be of the order of
$T^6$. In this paper, we present the black-body radiation formula in
$\kappa$-Minkowski space-time using the complex (free massless) scalar
field theory. It is to be noted that the scalar field theories
constructed in the $\kappa-$Minkowski space-time are not unique.
Depending on the ordering of the kernel of the Fourier transformation,
the multiplication law and the equations of motion for field operators
are differently realized~\cite{amelino2,LRZ}. And the different ordering
results in different dispersion relations for particles and
antiparticles. Here, we consider the ordering effect on the dispersion
relation of the particle and anti-particle spectrum. Another effect to
take into consideration is the measure in the momentum integral which
affects the counting of massless modes. We derive the black-body
radiation formula in two ways. Quantum mechanical approach is the one,
where the dispersion relation for the stable massless mode is used, and
an effective thermal field theory approach using  the free massless
scalar field theory is the other.

This paper is organized as follows.
In section \ref{stat}, we calculate the thermal energy density
based on the statistical mechanics of one particle quantum
mechanics in $\kappa-$Minkowski space-time.
In section \ref{sec:action}, we investigate the  ordering effect
on the free field action of a scalar field,
and second-quantize the theories.
The case of asymmetric
ordering is presented in subsection \ref{sec:asym-def} and the
symmetric ordering case in subsection \ref{sec:sym-def}. In the
asymmetric case the particle and antiparticle have different
dispersion relations, but there is a regularity in the energy
spectra of the complex modes. On the other hand, the symmetric
ordering has the same particle and antiparticle spectra but
the regularity in the energy spectra of the complex modes is lost.
In both cases, when the momentum reaches a certain value of order
$\kappa$, the energy dispersion
becomes infinite. When the momentum exceeds this value, then
some of complex modes become real modes and vice versa.
It turns out that the contribution of these high momentum modes to the
black-body radiation spectra does not have the correct commutative
limits.
In Section \ref{sec:thermalfieldtheory}, we propose an effective
thermal field theory in $\kappa$-Minkowski space-time using an
effective Hamiltonian, which reproduces the black-body radiation
spectra obtained in Section \ref{stat}. This is possible even in
this infinite derivative theory since the theory we are considering
is the diagonalized one and is decoupled from the infinite complex modes.
It is noted that in this formalism,
one naturally puts the
integration measure
as the $\kappa$-deformed Poincar\'e invariant one.
In section \ref{sec:summary}, our results are summarized
and some considerations on the high momentum mode
which exceeds the momentum cut-off are given.

%%%%%%%%%%%%%%%%%%%%%%%%%%%%%%%%%%%%%
\section{Black-body radiation}  \label{stat}
%%%%%%%%%%%%%%%%%%%%%%%%%%%%%%%%%%%%%
The dispersion relation of the massless particle
in $\kappa-$Minkowski space-time is known to
be different from that of the commutative one.
For example, in the case of the asymmetric deformation, the energies of
the stable modes with momentum $\bf k$ are given by
\begin{equation}
\label{eq:energy-dispersion} \Omega_{\bf k}^+= - \kappa \log(1-
\frac{|{\bf k}|}{\kappa})\,,\qquad \Omega_{\bf k}^-=  \kappa
\log(1+ \frac{|{\bf k}|}{\kappa}) \,.
\end{equation}
For the positive mode ( $\Omega_{\bf k}^+$), the momentum is
restricted to $|\bf k| < \kappa$. This dispersion relation is derived
from the asymmetric ordering of the kernel function of the Fourier
transformation (see the details in section \ref{sec:asym-def}).

These massless particles, in a thermal equilibrium at temperature
$1/\beta$, will give internal thermal energy density
\begin{equation}
\rho_+ (\beta)  = \int_{|{\bf k}|\leq\kappa} \frac{d^3 {\bf k}\,
e^{\alpha \Omega_{\bf
    k}^+/\kappa}}{(2\pi)^3}
    \frac{\Omega_{\bf k}^+}{e^{\beta \Omega_{\bf k}^+}-1}\,,\qquad
\rho_- (\beta) = \int \frac{d^3 {\bf k}\,e^{-\alpha \Omega_{\bf
    k}^-/\kappa}}{(2\pi)^3} \frac{\Omega_{\bf k}^-}{
    e^{\beta \Omega_{\bf k}^-}-1}\,,
\end{equation}
if one assumes that the particles do not interact with each other.
Here we introduce the measure factor $e^{\alpha
\Omega_{\bf k}^+/\kappa}\,$ ($e^{\alpha
-\Omega_{\bf k}^-/\kappa}$)
for the positive (negative) mode in the momentum integral.
This measure factor is  1 ({\it i.\/e.\/} $\alpha=0$)
if one simply counts the modes in momentum space.
On the other hand, if one requires the integration measure to be
invariant under the $\kappa-$Poincar\'{e} transformation, one needs to
set $\alpha=3$.
It is not clear at this point which one is the correct choice.
(More details are given in section \ref{sec:thermalfieldtheory}
and \ref{sec:summary}).

Explicit form of $\rho_+$ is
\begin{equation}
\rho_+ (\beta)  = \beta^{-4}  \frac{4\pi}{(2\pi)^3} f(\beta
\kappa)\,,\quad f(x)= x^4 \int_0^1 \, dp \frac{p^2 (1-p)^{-\alpha}\,
\log(1 -p)^{-1} } {(1-p)^{-x} -1}\,.
\end{equation}
One may expand $ f(x)$ as a series in a inverse power of  $x$ (by
setting $1-p = e^{-t}$),
\begin{align}
\label{fx} f(x)& =x^4 \int_0^\infty \, dt \frac{ t^3(a_0 + a_1 t + a_2
t^2 \cdots )} {e^{xt} -1} = \sum_{i=0}\Gamma(4+i)\zeta(4+i)
\frac{a_i}{x^i}
\nonumber\\
&= \frac{\pi^4}{15}+ \frac{ 24 \zeta(5)\, a_1 }{x}  + \frac{8\pi^6 \,
    a_2}{63\, x^2}+ \cdots\,.
\end{align}
The first few coefficients are given by $a_0=1, a_1=\alpha-2,~ a_2=
25/12-2\alpha+\alpha^2/2, ~a_3=-3/2
+25\alpha/12-\alpha^2+\alpha^3/6\, $ .

The negative mode  contribution to the energy density is given by
\begin{equation}
\rho_- (\beta) = \beta^{-4}  \frac{4\pi}{(2\pi)^3} g(\beta \kappa),\quad
g(x)= x^4 \int_0^\infty \, dp \frac{p^2(1+p)^{-\alpha}\, \log(1 +p) }
{(1+p)^{x} -1} .
\end{equation}
One may again expand $g(x)$ as a series in a inverse power of $x$
(by setting $1+p= e^{t}$),
\begin{align}
\label{gx}
g(x) &= x^4\int_0^{\infty} \,dt \frac{t^3( 1+ b_1 t + b_2 t^2 +
 \cdots )} {e^{xt} -1}
 = \sum_{i=0} \Gamma(4+i)\,\zeta(4+i)\, \frac{b_i}{x^i} \nn
\\
&=\frac{\pi^4}{15} + 24\zeta(5) \frac{b_1}x + \frac{8\pi^6}{63}
\frac{b_2}{x^2}+ \cdots.
\end{align}
where the coefficients are given by $b_1=2-\alpha,
~b_2=25/12-2\alpha+\alpha^2/2,~ b_3=3/2-
\frac{25}{12}\alpha+\alpha^2+\alpha^3/6\,$.
It is clear that the positive mode contribution
is not the same as the negative mode  contribution
since $b_i(\alpha)= (-1)^i a_i (\alpha)$ in Eqs.~(\ref{fx}) and
(\ref{gx}).

If the system consists of positive and negative modes together, then the
two contributions are added
\begin{align}
\label{eq:totaldensity}
\rho(\beta) &= \rho_+(\beta) +\rho_- (\beta) \\
&= \frac{T^{4}}{2 \pi^2}  \sum_{i=0} \Gamma(4+i)\,\zeta(4+i)\,
\frac{(a_{i} + b_i)}{x^{i}} \nn \\
&=  T^{4} \left( \frac{\pi^2}{15} + \frac{2 \pi^4}{189} \frac{(25-24\alpha+6\alpha^2)T^2}{\kappa^2}
    +  O\left(\frac {T^4}{\kappa^4} \right) \right) \,.\nn
\end{align}
This shows that to the lowest order, the internal energy density agrees
with the commutative result.
The  deviation from the commutative result comes at
$O(\frac{T^6}{\kappa^2})$.

The dispersion relation is different when derived from the
time-symmetric ordering of the kernel function of the Fourier
transformation as shown in Sec.~\ref{sec:sym-def}.
In this case the dispersion relations for the stable positive and
negative modes are the same and is given by $ \sinh \frac{\omega_{\bf
k}}{2\kappa}=\frac{|{\bf k}|}{2\kappa},$ or
\begin{equation}\label{wq:s-dispersion} \omega_{\bf k} = \kappa
    \log\left(1+\frac{{\bf k}^2}{2\kappa^2}+\frac{|{\bf
    k}|}{\kappa}\sqrt{1+\frac{\bf k^2}{4\kappa^2}}\right)\,.
\end{equation}

The thermal energy density for the positive mode is given
by
\begin{equation}
\rho_s^{(+)} (\beta) = \int \frac{d^3 {\bf k}\, e^{\bar\alpha \omega_{\bf
    k}/\kappa}}{(2\pi)^3}
    \frac{\omega_{\bf k}}{e^{\beta \omega_{\bf k}}-1}\,.
\end{equation}
The $\kappa-$Poincar\'{e} invariant measure factor is
achieved if one puts $\bar \alpha =3/2$.
The explicit expression for the energy density is given by
\begin{equation}
\rho_s ^{(+)}(\beta) = \beta^{-4}  \frac{4\pi}{(2\pi)^3} h(\beta \kappa)\,,
\qquad h(x)= x^4 \int_0^\infty \, dp \frac{p^2 A^{\bar \alpha}\, \log A
} {A^x -1},
\end{equation}
where $A =1+\frac{p^2}{2}+\sqrt{(1+p^2/2)^2-1}$.
Putting $A(p)= e^{t}$, one has
\begin{equation}
\label{hx}
h(x)= x^4\int_0^{\infty} \, dt
    \frac{ t  (1-e^{-t})^2\,e^{(\bar\alpha+1)t} \cosh(t/2)}
    {e^{xt} -1}
=x^4 \int_0^\infty  \, dt \frac{ t^3(c_0 + c_1 t
        + c_2 t^2\cdots )} {e^{xt} -1},
\end{equation}
where $c_0=1, c_1=\bar \alpha,~ c_2= 5/24+\bar\alpha^2/2, ~ c_3=5\bar
\alpha/24+\bar \alpha^3/6 $. Comparing $h(x)$ with $g(x)$ in
Eq.~(\ref{gx}), we have
\[
h(x)  =  \sum_{i=0}\Gamma(4+i)\zeta(4+i) \frac{c_i}{x^i}.
\]
Obviously, the negative mode contribution
$\rho_s ^{(-)}(\beta)$ is given if one replaces
$\bar\alpha$ to $-\bar \alpha$.
If the system has particle and anti-particle together, then the energy
density becomes
\begin{equation}
\label{eq:sym_density}
\rho_s ^{(+)}(\beta)+ \rho_s ^{(-)}(\beta)
= T^{4} \left(\frac{\pi^2}{15}
    + \frac{ \pi^4}{189} \frac{(5+12 \bar\alpha^2)T^2}{\kappa^2}
    +  O\left(\frac {T^4}{\kappa^4} \right) \right) \,.
\end{equation}
The energy density coincides with the commutative one at the
lowest order but the next order effect is not same as the
the previous asymmetric result Eq.~(\ref{eq:totaldensity}).
Thus, the ordering effect results in the
different blackbody radiation formula.

%%%%%%%%%%%%%%%%%%%%%%%%%%%%%%%%%%%%%%%%%%%%%%%%%%%%%%%%
\section{$\kappa$-deformed free field action}
\label{sec:action}
%%%%%%%%%%%%%%%%%%%%%%%%%%%%%%%%%%%%%%%%%%%%%%%%%%%%%%%%%%

The construction of field theories in $\kappa$-Minkowski space-time is
not simple since, for example, the
deformed Poincar\'{e} transformation
is not realized linearly in non-commutative coordinate
spacetime in so-called
bi-crossproduct basis ~\cite{majid}
(see Appendix for summary).
On the other hand, the $\kappa$-deformed Poincar\'e transformation can
be described confidently in the momentum space
as a dual description of the $\kappa-$Minkowski spacetime,
$\langle  p^\nu,x^\mu\rangle = i \eta^{\mu\nu}$ with
diag$(\eta^{\mu\nu}) =(1,-1,-1,-1)$.
Thus, we will construct the action for the scalar field theory in the
momentum space first and convert it into the one in coordinate space by
using the Fourier transformation.

The 4-momenta are required to commute with each other:
\begin{equation}\label{eq:comm-p}
\left[ p_\mu , p_\nu\right] = 0\,.
\end{equation}
In addition, the Lorentz algebra remains un-deformed from that of the
commutative case and the rotational invariance is maintained
\cite{majid}:
\begin{align}\label{eq:Lorentz-rotation}
& \left[ M_i, M_j \right] = i \epsilon_{ijk} M_k \,,
\quad
\left[ M_i, N_j \right] = i \epsilon_{ijk} N_k\,,
\quad
\left[ N_i, N_j \right] = -i \epsilon_{ijk} M_k
\\
& \left[ M_i, p_j  \right]
= i \epsilon_{ijk} p_k  \,,\qquad
\left[M_i, p_0 \right] = 0\,, \nn
\end{align}
where $ M_i $ and $N_i$ are rotation and boost generator respectively.

$\kappa$-deformed Poincar\'e algebra
deforms the commutators between the boost and the translation generators,
which reflects the non-commutative nature of the
space-time.
The explicit form of the commutators is not unique but
is related to the Hopf algebra of the co-product.

%%%%%%%%%%%%%%%%%%%%%%%%%%%%%%%%%%%%%%%%%%%%%%%%%%%%%%%
\subsection{Asymmetric deformation} \label{sec:asym-def}
%%%%%%%%%%%%%%%%%%%%%%%%%%%%%%%%%%%%%%%%%%%%%%%%%%%%%%%

The co-product of four momenta is given by~\cite{kosinski},
\begin{equation} \label{co-product:p}
\Delta(p^0)=p^0\otimes 1+1\otimes p^0\,, \qquad\qquad
\Delta(p^i)=p^i\otimes e^{-p^0/\kappa}+1\otimes p^i\,.
\end{equation}
This is related to a composition rule of the exponential operator $e^{-i
p \cdot \hat x}$, which is the kernel of the Fourier transformation and
transforms the theory in space-time coordinates to the one in momentum
space. Here $\hat x =(\hat x^0, \hat {\textbf{x}})$,
$p=(p^0,\textbf{p})$ and $p\cdot \hat x\equiv p^0 \hat x^0 - {\bf p
\cdot \hat x}$. Its ordering is defined as
\begin{eqnarray} \label{NOrdering}
:e^{-i p \cdot \hat x}\!\!:\, \equiv e^{-i p^0 \hat x^0}\, e^{i \bf p
\cdot \hat x}\,.
\end{eqnarray}
Multiplication of two ordered exponential functions follows from
Eqs.~(\ref{k-Minkowski}) and (\ref{NOrdering}):
\begin{eqnarray} %\label{multiplication}
  :e^{-i p \cdot \hat x}\!\!:\,:e^{-i q \cdot \hat x}\!\!: \,\, = \,\,:e^{-i (p^0+q^0)
      \hat x^0+ i ({\bf p}
     e^{-q^0/\kappa}+ {\bf q})\cdot{\bf \hat x}}\!\!:
\end{eqnarray}
whose four momentum addition rule indicates the
co-product~(\ref{co-product:p}). This co-product changes the commutation
relations between the boost  and translation generators:
\begin{align}\label{eq:boost}
&\left[ N_i, p_j\right] = i \delta_{ij} \Big( \frac
{\kappa}2 (1-e^{-2p^0/\kappa}) + \frac {{\bf p}^2} {2 \kappa} \Big)
-\frac{i}{\kappa} p_i p_j \,,\qquad \left[ N_i, p_0\right] =ip_i\,.
\end{align}

To construct field theory, one employs the $*$-product formalism where
all the coordinate variables $x$ commute with each other and the
homomorphism of the exponential non-commutative operator product is
maintained so that the non-commuting effect is encoded in the definition
of the $*$-product. From now on, instead of using the non-commutative
space-time coordinates directly, we will use the $*$-product formalism
of the exponential function defined as $e^{-i p \cdot  x}\ast e^{-i q
\cdot x} = e^{-i v (p, q) \cdot x }$ where
\begin{equation}
v (p, q) = ( p^0+q^0,
{\bf p} e^{-q^0/\kappa}+ {\bf q}) \,.
\end{equation}
An arbitrary function of coordinates is expressed
through the Fourier transformation of a momentum function,
\begin{equation}
F(x) = \int_p e^{-ip\cdot x }\, f(p)\,,
\end{equation}
where $\int_p$ denotes the integration over 4-momentum,
$\int d^4 p/(2\pi)^4$.

One needs $*$-derivatives of exponential function
to construct an action for a field theory.
The exterior derivative of deformed algebra is obtained in
\cite{kosinski,daszkie}:
\begin{equation}
\partial_\mu \ast e^{-ip x} \,= -i \chi_\mu (p) e^{-ip x}\,,
\end{equation}
where
\begin{align}\label{chi}
\chi_0(p) = \kappa\left[\sinh \frac{p_0}{\kappa}+\frac{{\bf
p}^2}{2\kappa^2} e^{p_0/\kappa}\right]\,,\qquad \chi_i(p)= p_i
e^{p_0/\kappa} \,.
\end{align}

The derivative function $\chi_\mu$ satisfies the simple
commutation relations with boost and rotation generators:
\begin{align}
\left[ N_i, \chi_j\right] = i \delta_{ij} \chi_0\,,\qquad \left[ N_i,
\chi_0\right] = i \chi_i \,,\qquad \left[ M_i, \chi_j\right] = i
\epsilon_{ijk} \chi_k \,,\qquad \left[ M_i, \chi_0\right] = 0\,.
\end{align}
Therefore, $\chi^\mu \chi_\mu (p) $ is $\kappa$-deformed
Poincar\'e invariant. The explicit expression for this function is
given by
\begin{equation}
\label{inv_chimu}
 \chi^\mu \chi_\mu (p)
 =M^2_\kappa(p)\left(1+\frac{M^2_\kappa(p)}{
    4\kappa^2}\right)
    \,,
\end{equation}
where $M_\kappa^2(p)$ is the first Casimir invariant ($\left[
M_{\mu\nu},M^2_\kappa(p) \right] =0$)
\begin{equation}
\label{Casimir} M_\kappa^2(p)=\left(2 \kappa \sinh \frac{p_0}{2 \kappa
}\right)^2- {\bf p}^2 e^{p_0/\kappa}\,.
\end{equation}
The invariant volume form $ d^4 \chi = d\chi_0 d\chi_1 d\chi_2 d\chi_3$
is related to the momentum integration;
 \begin{equation}
 d^4 \chi (p) = \frac {M^2_\kappa(p)}{2 \kappa^2} \, d^4 p\, e^{3p^0/\kappa} \,.
 \end{equation}
Thus  $d^4 p\, e^{3p^0/\kappa} $ is the invariant volume element.
Employing the notation $\tilde p$ where
\begin{equation}
\label{tilde-p}
 \tilde p^0 = p^0\,, \quad \tilde {\bf p} =  e^{p^0/\kappa}\,{\bf
 p}\,,
\end{equation}
the invariant measure is written as
\begin{equation}
d^4 p\, e^{3p^0/\kappa} = d^4 \tilde p\,,
\end{equation}
and its integration is denoted as $\int_{\tilde p} = \int d^4 \tilde p /(2\pi)^4 $.

The $*$-derivative has the conjugate counter-part
$\partial_\mu^\dagger$ through the defining relation
\begin{eqnarray}
\int_{x} \phi_1( x) *
 \Big( \partial_\mu \, \phi_2(\ x)\Big)
\equiv \int _{ x} \Big( \partial_\mu ^\dagger\, \phi_1( x)\Big) *  \phi_2(
x)\,. \nn
\end{eqnarray}
Its explicit representation is given by
\begin{equation}
\label{eq:conjugatederivative}
\partial_\mu^\dagger \,\ast \, e^{-ip \cdot x}
= -i  \chi_\mu  (-\tilde p ) \, e^{-ip \cdot x}\,.
\end{equation}
We will call $p \to - \tilde p$ the conjugate transformation
($\tilde p$ is given in Eq.~(\ref{tilde-p})). It
is noted that the  Casimir invariant $M^2_\kappa(p)$ and invariant
volume element do not change their form under the conjugate
transformation;
\begin{equation}
\label{eq:conjugate-symm}
 M_\kappa^2(p) = M_\kappa^2(-\tilde p )\,,\qquad
 d^4 \chi (p) = d^4 \chi (-\tilde p ).
\end{equation}

Before constructing the scalar field action, let us first consider the
equation of motion for a free scalar field. The classical field
equation will be given by
\begin{equation}
\label{eq:fieldequation}
 \left[\partial_\mu *  \,\partial^\mu* + m^2 \right] \, \phi(x) =0\,.
\end{equation}
Using the Fourier transformed form of the field
\begin{equation}
\label{eq:scalarfield} \phi(x) = \int_p e^{-ip\cdot x }\,
\varphi(p)\,,
\end{equation}
one finds the equation in momentum space ${\cal E} (p) \,
\varphi(p)=0\,$ where
\begin{equation}
{\cal E} (p)=  \chi^\mu \chi_\mu (p) -m^2
\end{equation}
is the $\kappa$-deformed invariant. The solution of
Eq.~(\ref{eq:fieldequation})  has the form in coordinate space
\begin{equation}
\label{eq:classicalsol} \Phi(x) = \int_p e^{-ip\cdot x }\, f(p)
\delta({\cal E} (p))\,,
\end{equation}
where $f(p)$ is an arbitrary  regular function.

It is obvious that the solution also satisfies the  conjugate equation
of motion,
\begin{equation}
\label{eq:conj-eq-motion}
 \left[\partial^\dagger_\mu *  \,
 \partial^{\dagger\mu} *
  + m^2 \right] \, \Phi (x) =0\,,
\end{equation}
due to the invariant property of $\chi_\mu$
in Eqs. (\ref{inv_chimu}) and
(\ref{eq:conjugate-symm}).
On the other hand,
noting \begin{equation} \label{exp-dagger}
\Big(:e^{-ip \hat x}:\Big)^\dagger = e^{-i \textbf{p} \cdot
\hat{\textbf{x} }} \,
  e^{i p^0 \hat x^0}
= \,\,:\! e^{i( p^0 \hat x^0 - e^{p_0/\kappa} \textbf{p}\cdot
 \hat{\textbf{x}})}\!\!:\,\,
= \,\, :\! e^{i \tilde p \cdot x }\!\!:
\end{equation}
one may define a conjugate scalar field of $\phi (x)$ in accordance with
the conjugation of the ordered exponential operator,
\begin{equation}
\label{eq:conju-field} \phi^{c}(x)\equiv \int_{\tilde p} \, e^{i \tilde
p \cdot  x}\,\varphi^\dagger(p) =\int_p e^{-ip \cdot
x}\,\,\varphi^\dagger(- \tilde p )\,.
\end{equation}
Here $\varphi^\dagger(p)$ denotes the ordinary complex conjugate of
$\varphi(p)$ in the momentum space and invariant measure $\int_{\tilde
p}$ is used.
Using $\phi^{c}(x)$ one may construct other solution of the equation of
motion
\begin{equation}
\Phi^c(x) =  \int_{\tilde p} e^{i \tilde p\cdot x }\,  f^\dagger(p)
\delta({\cal E} ( p)) \,.
\end{equation}
One concludes that if $\Phi(x)$ is the classical solution of the
equation of motion, then so is $\Phi^c(x)$. It is to be noted that if we
start with the exponential operator Eq.~(\ref{exp-dagger}) (with $p$
replaced by $-p$) instead of the original one in Eq.~(\ref{NOrdering}),
then the role of $\phi(x)$ and $\phi^{c}(x)$ are switched.

Thus, $\phi^{c}(x)$ is the analogue of the complex conjugate in
coordinate space.  To avoid the confusion with the ordinary
complex conjugate in commutative space, we put the superscript $c$
instead of $\dagger$.
The definition of the conjugation leads to the following
properties,
\begin{eqnarray}
\Big( \phi^c ( x) \Big)^c = \phi(x)\,, \qquad \Big( \phi_1 ( x) * \phi_2
(x) \Big)^c =  \phi_2^c  ( x) *\phi_1^c ( x)\,,\qquad \Big( \partial_\mu
\phi (x) \Big)^c = - \partial_\mu ^\dagger \, \phi^c (x)  \,.
\end{eqnarray}

It is natural to have the delta-function
\begin{eqnarray}
\int_p  e^{-ip\cdot x} = (2\pi)^4 \delta^4(x)
\end{eqnarray}
and its inverse $
\int_x  e^{-ip\cdot x} = (2\pi)^4 \delta^4(p) $
where $\int_x = \int d^4 x$.
Using this delta function, one has
\begin{align}
\label{inv-integra}
\int_{ x} \phi_1^c ( x) * \phi_2 (x) =  \int_{\tilde p}
\varphi_1^\dagger (p)\, \varphi_2 (p) =  \int_{p}  \varphi_1^\dagger
(-\tilde p)\, \varphi_2 (-\tilde p)\,,
\end{align}
which is invariant under the deformed Poincar\'e transformation if
$\varphi_1^\dagger (p)\, \varphi_2 (p)$ is.
It should be noted that
the Poincar\'e transformation property of the space-time coordinates
is not defined clearly, but one can always check if
the integrated value is Poincar\'e invariant in momentum space.

The classical equation can be obtained from  action,
\begin{align}
S= \int_x \, \phi^c(x)*
 \left[-\partial_\mu *  \,
 \partial^\mu*
  - m^2 \right] \, \phi(x)
 = \int_{\tilde p} \, \varphi^\dagger(p) \Delta_F ^{-1}(p)
 \, \varphi(p) \,,
\end{align}
where
\begin{equation}
\label{eq:feynmanprop} \Delta_F ^{-1}(p) = {\cal E}(p) + i\epsilon
\end{equation}
will be called the Feynman propagator and  $\epsilon$, the small
positive real number, is added to avoid the singularity on the
real axis of $p^0$. The scalar fields in coordinate space,
$\phi(x)$ and $\phi^c(x)$ are defined in
Eq.~(\ref{eq:scalarfield}) and (\ref{eq:conju-field}). It is noted
that the action is $\kappa$-Poincar\'e invariant if
$\varphi^\dagger(p) \varphi(p) $ is. Thus we will require $
\varphi(p) $ and $\varphi^\dagger(p)  $ be $\kappa$-Poincar\'e
invariant.

The Feynman propagator is factorized as $\Delta_F ^{-1}(p) = F_+(p)
F_-(p) $ where
\begin{align}\label{eq:deltaF-factorization}
&F_+(p) =  \frac{\kappa}2 e^{p_0/\kappa}
    (e^{-p_0/\kappa} + \alpha_+)(e^{-p_0/\kappa} + \alpha_-) \,,
    \nn \\
&F_-(p) =  \frac{\kappa}2 e^{p_0/\kappa}
    (e^{-p_0/\kappa} - \alpha_+)(e^{-p_0/\kappa} - \alpha_-) \,,
\end{align}
with
\[
\alpha_\pm  =
 \sqrt{1+ \frac{m^2-i\epsilon}{\kappa^2}}
\pm \sqrt{ \frac{m^2+\textbf{p}^2 -i\epsilon}{\kappa^2}}\,.
\]
$F_\pm (p) $ is invariant under the conjugate transformation $
F_\pm (p) =F_\pm (- \tilde p )\,$.
It is noted that
$\Delta_F ^{-1}(p)$ is periodic in $p_0$ with a period $i \kappa \pi$;
\begin{equation}
\label{period-propagator} \Delta_F ^{-1}(p_0+ i \kappa \pi, {\bf p} )
=\Delta_F ^{-1}(p_0, {\bf p})\,.
\end{equation}
Thus, the on-shell condition is given by
\begin{equation}
p_0 = \Omega^{+}_{\bf p} + i n \pi \kappa \,,\,\,\,
-\Omega^{-}_{\bf p} + i n \pi \kappa
\end{equation}
where $n$ is an integer and
\begin{equation}
\Omega^{+}_{\bf p} = - \kappa\ln \alpha_- \,,\qquad
\Omega^{-}_{\bf p} = + \kappa\ln \alpha_+ \,.
\end{equation}
The existence of the
infinite number of on-shell values is due to the infinite number
of time derivatives in the $*$-equation of motion.

Here $\Omega_{\bf p}^- $ is always positive for real $\textbf{p}$, but
$\Omega_{\bf p}^+$ is positive only when $ \textbf{p}^2 < {\kappa^2}$.
If $ \textbf{p}^2 > {\kappa^2}$, $\Omega_{\bf p}^+$ becomes complex but
$-\kappa\ln (-\alpha_ -)$ becomes real. The meaning of  this mode change
is not clear at this moment. In this section we proceed to
second-quantize  the  positive and negative on-shell modes of this
scalar field using the momentum cut-off. Some more consideration is
given in section \ref{sec:summary}.

Suppose one try to second-quantize the field.
To do this one may introduce the source terms to the action
\[
\int_x   \Big( J^c(x) \phi(x)
+  \phi^c(x) J(x) \Big)
=
\int_{\tilde p} \Big( j^\dagger(p) \varphi(p)
+  \varphi^\dagger(p) j(p) \Big)
\]
and identify the field correlation as
\[
\left\langle \varphi(p)
\varphi^\dagger (q) \right\rangle \, e^{3q^0/\kappa}
=\left\langle \varphi^\dagger (q)
\varphi(p) \right\rangle \,  e^{3q^0/\kappa}
= i\Delta_F(p) (2\pi)^4 \,\delta^4 (p-q)\,.
\]
In analogy with the commutative field theory,
one may introduce the second-quantized field operator
with a prescription when integrated over the momentum space.
Putting
$\Phi_{\text {op}} ({\bf p})
= \Phi_{\text {op}}^{(+)} ({\bf p})
+\Phi_{\text {op}}^{(-)} ({\bf p})$
(superscript $+/-$  refers to the
positive/negative frequency part)
\[
\left[ \Phi_{\text{op}}^{(+)} ({\bf p}),
\Phi_{\text {op}}^{\dagger(-)} ({\bf q}) \right]
= \left \langle \Phi_{\text {op}}^{(+)} ({\bf p})
\Phi_{\text {op}}^{\dagger(-)} ({\bf q}) \right\rangle
= i \oint_{\text{lower}} \frac{dp^0}{2\pi} \Delta_F(p)
(2\pi)^3 \,\delta^3 ({\bf p}-{\bf q})\,.
\]
$\oint_{\text{lower}}$ is the prescription
of the integral of $p^0$ so that
the contour is taken along the lower half-plane
of the complex $p^0$ plane
to include the real pole $\Omega_{\bf p}^+$.
Explicit evaluation results in the quantization rule,
\begin{equation}
\label{comm-phi+}
\left[ \Phi_{\text{op}}^{(+)} ({\bf p}),
\Phi_{\text {op}}^{\dagger(-)} ({\bf q}) \right]
=  \sum_{\text{mode}} \frac{(2\pi)^3 \,\delta^3 ({\bf p}-{\bf q})}
{ 2 E ({\bf p})} \,.
\end{equation}
$E({\bf p})= \sqrt{(m^2 + {\bf p}^2)(1+ m^2/\kappa^2)}$
comes from the residue
\[{ \left| \frac{\delta  \Delta_F^{-1} (p)}{\delta p^0}
\right|}_{p^0= \Omega_{\bf p}^+ \pm i n \pi \kappa  } = {2E({\bf p})}\,.
\]
Note that the summation in (\ref{comm-phi+})
is due to the periodic properties of the pole structure in the Feynman propagator (\ref{period-propagator}).
Each pole in the complex plane contributes to each mode sum.
Likewise,
\begin{eqnarray}
\left[ \Phi_{\text{op}}^{(-)} ({\bf p}),
\Phi_{\text {op}}^{\dagger(+)} ({\bf q}) \right]
& =& - \left \langle
\Phi_{\text {op}}^{\dagger(+)} ({\bf q})
\Phi_{\text {op}}^{(+)} ({\bf p})
 \right\rangle
= -i \oint_{\text{upper}} \frac{dp^0}{2\pi} \Delta_F(p)
(2\pi)^3 \,\delta^3 ({\bf p}-{\bf q})
\nn
\\
&= & \sum_{\text{mode}} \frac{(2\pi)^3 \,\delta^3 ({\bf p}-{\bf q})}
{ 2 E ({\bf p})} \,.
\end{eqnarray}
$\oint_{\text{upper}}$ is the contour taken
along the upper half-plane to include the pole
$-\Omega_{\bf p}^-$.
This identification results in the second-quantized field
\begin{align*}
\Phi_{\text {op}}^{(+)} ({\bf p})
&= \frac1{ 2 E ({\bf p})} \Big(
a( {\bf p})  +
  \sum_{n>0} ( a_n( {\bf p})
    + b_n( -{\bf p}) ) \Big)
\\
\Phi_{\text {op}}^{(-)} ({\bf p})
&= \frac1{ 2 E ({\bf p})} \Big(
b^\dagger( {\bf p})  +
  \sum_{n<0} (a_n^\dagger( {\bf p})
  + b_n^\dagger(-{\bf p}))\Big)
\\
\Phi_{\text {op}}^{\dagger(+)} ({\bf p})
&= \frac1{ 2 E ({\bf p})} \Big(
a^\dagger( {\bf p})  +
  \sum_{n>0} ( a_n^\dagger( {\bf p})
    + b_n^\dagger( -{\bf p}) ) \Big)
\\
\Phi_{\text {op}}^{\dagger(-)} ({\bf p})
&= \frac1{ 2 E ({\bf p})} \Big(
b {\bf p})  +
  \sum_{n<0} (a_n( {\bf p}) + b_n(-{\bf p}))\Big)
\end{align*}
where the creation and annihilation operators satisfy the commutation relations
\begin{align}
\left[ a( {\bf p}), a^\dagger ({\bf q}) \right]
 &=2 E({\bf p})\,
 (2\pi)^3  \delta^3( {\bf p} -{\bf q}) \,,\quad
 \left[ a_n( {\bf p}), a_m^\dagger ({\bf q}) \right]
 =2 E({\bf p})\,
 ( 2\pi)^3  \delta^3( {\bf p}-{\bf q}) )\,\delta_{m,n} \,,
 \nn
 \\
\label{comm}
 \left[ b( {\bf p}), b^\dagger ({\bf q}) \right]
 &=2 E({\bf p})\,
 (2\pi)^3  \delta^3( {\bf p}-{\bf q})  \,,\quad
 \left[ b_n( {\bf p}), b_m^\dagger ({\bf q}) \right]
 =2 E({\bf p})\,
  (2\pi)^3  \delta^3( {\bf p}-{\bf q}) )\,\delta_{m,n} \,.
\end{align}

The quantum field in coordinate space may be defined as
\begin{align} \label{eq:quantumscalarfield}
\Phi_{\text{op}}(x) &=\int
\frac{d^3 {\bf p}}{ (2\pi)^3
2E({\bf p})}\,
  \left[  {e^{-ip_+\cdot x} }\,  a( {\bf p})  +
  \sum_{n>0}  e^{-n\pi \kappa | x^0|}
     {e^{-ip_+\cdot x} } \,
    \Big(a_n( {\bf p}) +a_{-n}^\dagger( {\bf p}) \Big)
     \right]\,\\
 & +\int \frac{d^3 {\bf p} }
                    { (2\pi)^3 2E({\bf p})}\,
  \left[
    {e^{i \tilde p_-\cdot x} }\, b^\dagger( {\bf p})
    + \sum_{n>0}  e^{-n\pi \kappa | x^0|}
    {e^{i \tilde p_-\cdot x} } \,
    \Big( b_n( {\bf p})     + b_{-n}^\dagger ( {\bf p}) \Big)
     \right]\, , \nn \\
\Phi_{\text{op}}^c(x) &=\int %_{|{\bf p}|\leq \kappa}
\frac{d^3 {\bf p}\, e^{3\Omega^+_{\bf p}/\kappa}}{ (2\pi)^3
    2E({\bf p})}\,
  \left[ {e^{i\tilde p_+\cdot x} }\,  a^\dagger( {\bf p})  +
  \sum_{n>0}  e^{-n\pi \kappa | x^0|}
     {e^{i\tilde p_+\cdot x} } \,
    \Big(e^{3i n\pi/\kappa}a_n^\dagger( {\bf p})
    + e^{-3i n\pi/\kappa} a_{-n}( {\bf p}) \Big)
     \right]\,\nn \\
  +&\int \frac{d^3 {\bf p}\, e^{-3\Omega^-_{\bf p}/\kappa}}
        { (2\pi)^3 2E({\bf p})}\,
  \left[
    {e^{-i p_-\cdot x} }\, b( {\bf p})
    + \sum_{n>0}  e^{-n\pi \kappa | x^0|}
    {e^{-i p_-\cdot x} } \,
    \Big(e^{3i n\pi/\kappa} b_n^\dagger( {\bf p})
    + e^{-3i n\pi/\kappa} b_{-n} ( {\bf p}) \Big)
     \right]\, ,\nn
\end{align}
where $p_+ = ( \Omega_{\bf p}^+, {\bf p})$ and
$p_- = ( \Omega_{\bf p}^-, {\bf p})$.
This quantized field satisfies the equation of motion
(\ref{eq:fieldequation}) and its conjugate one.
$a^\dagger ({\bf p}) $ creates a positive mode of energy $\Omega^+_{\bf
p}$ and momentum ${\bf p}$ from the vacuum $|0\rangle$ with momentum
$|\bf p|< \kappa$. (The momentum integration of $a^\dagger ({\bf p}) $
and $a({\bf p}) $ mode in (\ref{eq:quantumscalarfield}) is assumed to be
limited by $\kappa$). Likewise, $b^\dagger ({\bf p}) $ creates a
negative mode of energy $\Omega^-_{\bf p}$ and momentum ${\bf p}$. (One
may identify the positive mode with a particle and the negative mode
with an anti-particle or vice versa.)

It should be remarked that the quantum field theory
in $\kappa$-Minkowski spacetime is not well established yet,
even though the above procedure is one of plausible proposals.
For example, the complex mode with nonzero imaginary
energy comes from the periodicity of the
Feynman propagator Eq.~(\ref{period-propagator})
which originates from the non-trivial form of the Casimir invariant
Eq.~(\ref{inv_chimu}).
The presence of the complex mode is not of the usual form of the
canonical field theory
and may raise the unitarity problem.
Nevertheless, the complex mode
lives only for a short period of time of the order $1/(n\kappa)$
and for this reason, we do not include the complex modes
in the calculation of the black body radiation in Sec. II.
In a similar spirit, one may find some
authors neglect this complex modes from the beginning
\cite{realmodefieldtheory}.
Another concern is that the
relation between the propagator in coordinate space
and one in momentum space
is not clearly defined yet.
One might obtain the Feynman propagator
in coordinate space directly from the given action.
In this process, however, one needs to evaluate
the functional derivative inside the $*$-product and
also define an appropriate $*$-product
between different spacetime points.
This non-trivial questions are to be
treated carefully for a consistent
and complete field theory,
whose analysis is beyond the scope of our paper.
This issue will be treated in a seperate paper
along with interacting theory.

Let us turn to the real field case.
The dispersion relations of the positive mode
and negative mode energy parts do not coincide when ${\bf p} \ne 0$:
\begin{equation}
 \Omega^+_{\bf p} - \Omega^-_{\bf p} =
  -\kappa\ln (\alpha_+ \alpha_-) =
-\kappa\ln ( 1- \textbf{p}^2/\kappa^2 ) \,,
\end{equation}
which is positive when $|\textbf{p} | < \kappa$.
This mismatch leads to the difficulty in defining the real field as a
self conjugate operator. As a consequence, one may use either of
the following expansions as a candidate for real field
representation:
\begin{align}
\label{eq:realfield} \Phi_1(x) &=\int \frac{d^3 p}{ (2\pi)^3 2 E({\bf
p})}\,
  \left[  {e^{-ip_+\cdot x} } a( {\bf p}) +
   \sum_{n>0}  e^{-n\pi \kappa | x^0|}
    {e^{-ip_+\cdot x} }  \Big(a_n( {\bf p}) +a_{-n}^\dagger( {\bf p})
    \Big) \right] \nn\\
&+ \int \frac{d^3 p}
    {(2\pi)^3  2 E ({\bf  p})} \, \left[
     {e^{i \tilde p_+\cdot x} }
    a^\dagger( {\bf p})  +
     \sum_{n>0}  e^{-n\pi \kappa |x^0|} {e^{i\tilde p_+ \cdot x} }
  \Big(a_n^\dagger( { \bf p}) +a_{-n}( {\bf p})\Big) \right],\\
\Phi_2(x) &=\int \frac{d^3 p}{ (2\pi)^3 2 E({\bf p})}\,
  \left[  {e^{-i p_-\cdot x} } b( {\bf p}) +
   \sum_{n>0}  e^{-n\pi \kappa | x^0|}
    {e^{-ip_-\cdot x} }  \Big(b_n( {\bf p}) +b_{-n}^\dagger( {\bf p})
    \Big) \right] \nn\\
&+ \int \frac{d^3 p}
    {(2\pi)^3  2 E ({\bf  p})} \, \left[
     {e^{i \tilde p_-\cdot x} } b^\dagger( {\bf p})  +
     \sum_{n>0}  e^{-n\pi \kappa |x^0|} {e^{i\tilde p_- \cdot x} }
  \Big(b_n^\dagger( { \bf p}) +b_{-n}( {\bf p})\Big) \right]\,.
\end{align}
This raises a question: Can we choose a vacuum or an energy operator
such that the positive and negative spectra are symmetric? Or is it
possible to shift the on-shell value $ p^0 \to p^0 - (\Omega^+_{\bf p} -
\Omega^-_{\bf p})/2$, and redefine the on-shell value so that positive
and negative dispersion relations are the same, $ \tilde \Omega^+_{\bf
p} = \tilde \Omega^-_{\bf p} = \frac{\kappa}2 \ln \left({\alpha_+
}/{\alpha_- } \right)\,$? To answer this question, we use a freedom to
choose the definition of the $*$-product by redefining the ordering of
the exponential function in the following subsection.

%%%%%%%%%%%%%%%%%%%%%%%%%%%%%%%%%%%%%%%%%%%%%%%%%%
\subsection{Symmetric deformation} \label{sec:sym-def}
%%%%%%%%%%%%%%%%%%%%%%%%%%%%%%%%%%%%%%%%%%%%%%%%%%

Let us newly define the ordering of exponential function in a symmetric
way~\cite{amelino2}:
\begin{equation}
 \left[e^{-ip \hat x}\right]_s  \equiv e^{-ip^0\hat x^0/2}
 e^{i{\bf p} \cdot \hat {\bf x} }
 e^{-ip^0\hat x^0/2}\,.
\end{equation}
The multiplication of the two symmetric exponential functions is
given by
\begin{equation}
\left[e^{-i p  \hat x}\right]_s \, \left[e^{-i q \hat x}\right]_s =
\left[ e^{ -i (p^0+q^0)\hat x^0 + i ({\bf p} e^{- \frac{q^0}{2\kappa}}+
{\bf q} e^{ \frac{p^0}{2\kappa}}) {\bf x}}\right]_s \,.
\end{equation}
Therefore, the new multiplication rule is given as $e^{-i p  x}*_s e^{-i
q x} = e^{- i v(p,q)x}$ where
\begin{equation}
  v(p,q) = (p^0+q^0, {\bf p} e^{- \frac{q^0}{2\kappa}}+ {\bf q} e^{
  \frac{p^0}{2\kappa}})\,.
\end{equation}
This is equivalent to changing the co-product of $p_\mu$ as
\begin{equation}
\Delta(p^0) = 1 \otimes p^0 + p^0 \otimes 1\,,\qquad \Delta(p^i) =
p^i \otimes e^{\frac{-p^0}{2\kappa}} +e^{\frac{p^0}{2\kappa}}
\otimes p^i \,,
\end{equation}
and the commutation relation involving the boost generator in
$\kappa-$deformed Poincar\'e algebra as
\begin{align}
\label{eq:s-poincare} \left[ N_i, p_j\right] = i \delta_{ij} \,
e^{\frac{p^0}{2\kappa}} \Big( \frac {\kappa}2 (1-e^{-2p^0/\kappa}) +
\frac {{\bf p}^2} {2 \kappa}  e^{-\frac{p^0}{\kappa}} \Big)
-\frac{i}{2\kappa} p_i p_j  e^{-\frac{p^0}{2\kappa}} \,,\qquad \left[
N_i, p_0\right] =ip_i e^{-\frac{p^0}{2\kappa}} \,.
\end{align}
Other algebraic relations involving  translation and rotation are not
affected. This change is effectively summarized as the shift of the
3-momenta ${\bf p} \to {\bf p} e^{-\frac{p^0}{2\kappa}}$ in the
asymmetric definition, Eq.~(\ref{co-product:p}) and (\ref{eq:boost}).

The $*_s$-derivative of exponential function now takes the form
\begin{align}\label{xi}
&\partial_\mu*_s e^{-ip x} \,= -i \xi_\mu (p) e^{-ip x}\, ,
\\
& \xi_0(p) = \kappa\left[\sinh \frac{p_0}{\kappa}+\frac{{\bf
p}^2}{2\kappa^2}\right]\,,\qquad \xi_i(p)= p_i e^{\frac{p_0}{2\kappa}}
\,, \nn
\end{align}
and its conjugate partial derivative is simply given by
\begin{equation}
\label{eq:s-conjugatederivative}
\partial_\mu^\dagger \, *_s e^{-ip \cdot x}
= -i  \xi_\mu  (- p ) \, e^{-ip \cdot x}\,.
\end{equation}
Under the conjugate transformation, $p$ simply changes to $-p$.

The Poincar\'{e} symmetry of the derivative functions $\xi_\mu$ remains
the same;
\[
\left[ N_i, \xi_j\right] = i \delta_{ij} \xi_0\,,\qquad \left[ N_i,
\xi_0\right] = i \xi_i \,,\qquad
 \left[ M_i, \xi_j\right] = i \epsilon_{ijk} \xi_k \,,\qquad
\left[ M_i, \xi_0\right] = 0\,,
\]
and, as in the case of the asymmetric ordering, $\xi^\mu \xi_\mu
(p) $ is invariant. The explicit expression for the invariant is
\begin{equation}
 \xi^\mu \xi_\mu (p) = M^2_s(p)\left(1+\frac{M^2_s (p)}{
    4\kappa^2}\right)
    =\xi^\mu\,\xi_\mu (p) = \xi^\mu\,\xi_\mu (-p)\,,
\end{equation}
where the  Casimir invariant  $M_s^2(p)$ has a different form:
\begin{equation}
M_s^2(p)=\left(2 \kappa \sinh \frac{p_0}{2 \kappa }\right)^2- {\bf p}^2
= M_s^2(-p)\,.
\end{equation}
In addition, the invariant  volume form $d^4 \xi = d\xi_0 d\xi_1
d\xi_2 d\xi_3$ is given in terms of the momentum variables by
 \begin{equation}
 d^4 \xi (p) = \frac {M^2_s(p)}{2 \kappa^2} \, d^4 p\,
 e^{\frac {3p^0}{2\kappa} } \,.
 \end{equation}
Thus the invariant measure for the momentum integral becomes
\begin{equation}
d^4 p_s \equiv d^4 p\,  e^{\frac {3p^0}{2\kappa} }\,.
\end{equation}

Now the equation of motion for a free scalar field will be given
by
\begin{equation}
\label{eq:s-fieldequation}
 \left[\partial_\mu *_s  \,\partial^\mu*_s + m^2 \right] \, \phi_s(x) =0\,.
\end{equation}
One may use the Fourier transformed field
\begin{equation}
\phi_s(x) = \int_{p} e^{-ip\cdot x }\, \varphi(p)\,.
\end{equation}
The equation of motion becomes in momentum space
\begin{equation}
{\cal H} (p) \, \varphi(p)=0\,,\qquad {\cal H} (p)=  \xi^\mu \xi_\mu (p)
-m^2
\end{equation}
and ${\cal H} (p)$ is the $\kappa-$deformed Poincar\'{e} invariant.

The solution of the classical equation of motion,
Eq.~(\ref{eq:s-fieldequation}),  has the form
\begin{equation}
\label{eq:s-classicalsol} \Phi_s(x)
= \int_{p} e^{-ip\cdot x }\, f(p)
\delta({\cal H} (p))\,,
\end{equation}
with $f(p)$ an arbitrary  regular function.
This solution also satisfies the conjugate equation,
\begin{equation}
 \left[\partial^\dagger_\mu *  \,
 \partial^{\dagger\mu} *
  + m^2 \right] \, \Phi_s (x) =0\,.
\end{equation}

On the other hand, using the complex conjugation property
\begin{equation} \label{s-exp-dagger}
\Big[e^{-ip \hat x} \Big]^\dagger = \Big[ e^{i  p \cdot \hat x }
\Big]\,,
\end{equation}
one may define the conjugate of $\phi_s(x)$ as
\begin{equation}
\phi^{c}_{s}(x) =  \int_{ p} e^{i p\cdot x }\, \phi^\dagger (p)\,,
\end{equation}
just the complex conjugate of $\phi_{s}(x) $
Thus, $\Phi^{c}_{s}(x) =  \int_{ p} e^{i p\cdot x }\, f^\dagger (p)
\delta({\cal H} ( p))$
is again a solution of the conjugate equation of
motion since
\begin{equation}
\label{eq:s-conjugate-fieldequation}
\partial_\mu^\dagger \, \, e^{i  p \cdot x}
= -i  \xi_\mu  (p ) \, e^{i  p \cdot x}\,,
\end{equation}
and  ${\cal H} (- p) = {\cal H} ( p)$. This is similar to the ordinary
complex conjugation except for the change of measure. It should be noted
that
\begin{equation}
\int_x \phi_s^c (x) * \psi_s(x)
=\int_{p_s} \phi^\dagger(p) \psi(p)
\end{equation}
is $\kappa$-deformed Poincar\'e invariant if
the Fourier-transformed
$ \phi^\dagger(p) \psi(p) $ is
since $\int_{p_s}\equiv \int d^4 p_s/(2\pi)^4 $
is invariant.

The equation of motion can be obtained from the
$\kappa$-deformed Poincar\'e invariant action,
\begin{align}
S_2&= \int_x \, \phi^{c}_{s} (x)*_s
 \left[-\partial_\mu *_s  \,
 \partial^\mu*_s
  - m^2 \right] \, \phi_s(x)
\nn \\
&= \int_{p_s} \, \varphi^\dagger(p) \Delta_S ^{-1}(p)
 \, \varphi(p)
\end{align}
where $\Delta_S ^{-1}(p)$ is modified as
\begin{equation} \label{eq:propagator:S}
\Delta_S ^{-1}(p) = {\cal H}(p) +  i\epsilon =\Delta_S ^{-1}(-p)
\end{equation}
with $\epsilon$ small real number to avoid the singularity on the
real axis of $p^0$.

The one-shell condition ($\Delta_S^{-1}(p) =0$) is  given by
\[
\cosh\frac{p^0}{\kappa} = \frac{\textbf{p}^2}{2\kappa^2} \pm  \sqrt{1+
\frac{m^2-i\epsilon}{\kappa^2}} \,.
\]
The case for the upper (+) sign  yields the two  real solutions
$\pm \omega_{\bf p} $, where $\omega_{\bf p} = \kappa \ln \beta
>0$:
\begin{equation}
\beta =a + \sqrt{a^2 -1}\,,\qquad a = \frac{\textbf{p}^2}{2\kappa^2}+
\sqrt{1+ \frac{m^2-i\epsilon}{\kappa^2}}\,.
\end{equation}
In fact, $\Delta_S ^{-1}(p)  $ is factorized as
\begin{align}\label{eq:DeltaS}
\Delta_S ^{-1}(p) = \frac{\kappa^2}4 e^{2p_0/\kappa}
    (e^{-p_0/\kappa} - \beta)(e^{-p_0/\kappa}  - \frac1\beta)
    (e^{-p_0/\kappa} - \gamma)(e^{-p_0/\kappa} -\frac1\gamma)
\end{align}
where
\[
\gamma =b + \sqrt{b^2 -1} \,,\qquad b =\frac{\textbf{p}^2}{2\kappa^2} -
\sqrt{1+ \frac{m^2-i\epsilon}{\kappa^2}}
\]
and has the periodicity
\begin{align}
\label{eq:periodicity-deltas} \Delta_S ^{-1}(p) &=\Delta_S ^{-1} (p_0+ 2
i \kappa \pi )\,.
\end{align}
There is no simple relation between $\beta$ and $\gamma$ as in the
asymmetric case, Eq.~(\ref{eq:deltaF-factorization}): The period of
$\Delta_S ^{-1}(p)$ is $2 i \kappa \pi$ rather than $ i \kappa \pi$.  In
addition, there exists a real and complex mode change
as the momentum $\textbf{p}$
exceeds certain value of order of $\kappa$, which appears in the
asymmetric case also: $\gamma $ is complex when $ {\textbf{p}^2}$ is
small
\begin{equation}
\label{eq:momentumcutoff}
\frac{\textbf{p}^2}{2\kappa^2} < 1+ \sqrt{1+
\frac{m^2-i\epsilon}{\kappa^2}}\,,
\end{equation}
and its on-shell value of the energy $\pm \kappa (\ln \gamma)$ is complex.
However, when $ {\textbf{p}^2}$ exceeds
the above limit Eq.~(\ref{eq:momentumcutoff}),
the pair of the complex valued energies  becomes
a pair of real positive and negative energies.

Second-quantized scalar field can be obtained similarly as in the
asymmetric case in Eq.~(\ref{eq:quantumscalarfield}).
However, the
residue for the $p^0$ integration is not simple as in the
asymmetric case due to the periodicity in
Eq.~(\ref{eq:periodicity-deltas}). Nevertheless, defining the
residue at $\omega_{\bf p} $ as
\begin{align*}
D({\bf p})= \left| \frac{\delta  \Delta_S^{-1} (p)}{\delta p^0}
\right|_{p^0= \pm \omega_{\bf p}}
\end{align*}
one may write the scalar field  as
\begin{align}
\label{eq:s-quantumscalarfield} \Phi_s(x) &=\int \frac{d^3p}{(2\pi)^3\,
2D({\bf p})} \,
  \Big( {e^{-ip \cdot x} }\,  a( {\bf p}) +
   {e^{ip\cdot x} }\,  b^\dagger( {\bf p})\Big)
    + \cdots \,, \\
\Phi^{c}_s  (x) &= \int \frac{d^3p}{(2\pi)^3 \, 2D({\bf p})} \,
    \Big( e^{\frac{3 \omega_{\bf p}}{2\kappa}} {e^{i p \cdot x} }\,  a^\dagger( {\bf p})
    +   e^{-\frac{3 \omega_{\bf p}}{2\kappa}}{e^{-i p \cdot x} }\,  b( {\bf p}) \Big)
   + \cdots \nn\,,
\end{align}
where  $\cdots$ denotes the contributions from the complex modes
and higher energy modes at the Planck scale
and the momentum is limited as in Eq.~(\ref{eq:momentumcutoff}).
The commutation relations become
\begin{align}
\label{s-comm} \left[ a( {\bf p}), a^\dagger ({\bf q}) \right]
 =2 D({\bf q})  \, (2\pi)^3  \delta^3( {\bf p}-{\bf q}) \,,\qquad
 \left[ b( {\bf p}), b^\dagger ({\bf q}) \right]
 =2 D({\bf q})\,  (2\pi)^3  \delta^3( {\bf p}-{\bf q}) \,.
\end{align}
Both
$a^\dagger ({\bf p}) $ and
$b^\dagger ({\bf p}) $  create modes with energy $\omega_{\bf p}$ and momentum ${\bf p}$.
This shows that the real field can be
self-conjugate if one demands
\begin{equation}
a({\bf p}) = b({\bf p}) \quad
\textrm{or}\quad
\Phi_s^c(x) = \Phi_s (x) \,.
\end{equation}

%%%%%%%%%%%%%%%%%%%%%%%%%%%%%%%%%%%%%%%%%%%%%%%%%%%
\section{Effective thermal field theory}
\label{sec:thermalfieldtheory}
%%%%%%%%%%%%%%%%%%%%%%%%%%%%%%%%%%%%%%%%%%%%%%%%

The free scalar field theory constructed in $\kappa$-Minkowski
space-time in the last section leads to the massless dispersion relation
$\pm \Omega_{\bf k}^\pm$ of Eq.~(2) in the asymmetric ordering case, and
$\pm\omega_{\bf k}$ of Eq.~(9) in the symmetric ordering case. Based on
these dispersion relations, we have used the quantum mechanical particle
picture to evaluate the black-body radiation spectra for the massless
scalar particles living in the
$\kappa$-Minkowski space-time.%

In the case of the conventional commutative field theories, the
black-body radiation spectrum evaluated by quantum mechanical
method, agrees with that computed by using the thermal field
theory, which is formulated based on the observation that the time
evolution operator in quantum field theory can be analytically
continued to the Boltzmann factor of the quantum statistical
mechanics.

However, the field theory in the $\kappa$-Minkowski space-time involves
infinite order time derivatives and one can not define the Hamiltonian
operator as in the conventional field theory.
Due to this difficulty  we do not have a reliable field theoretic method
to evaluate the quantum partition function.
Nevertheless, for this diagonalized theory of the scalar field considered
in the last section, one can proceed with the individual modes. Note
that there are two stable modes $\pm\Omega^\pm_{\bf k}$ in the
asymmetric case and $\pm\omega_{\bf k}$ in the symmetric case. If there
are two stable modes only, we may start with an effective ``Hamiltonian"
\begin{align}\label{eq:hamiltonian}
H[\phi] &= \int_x  \phi^\dagger \Big(\frac{\partial} {\partial \tau} +
\tilde \omega^+(-i\nabla)\Big) \Big(\frac{\partial} {\partial \tau} +
\tilde \omega^-(-i\nabla) \Big) \phi(x)
\end{align}
with Euclidean ``time" $\tau$. $\tilde \omega^\pm(-i\nabla)$ reduces to
$\tilde \omega^\pm ({\bf k})$ in momentum space and refers to the
$\Omega^\pm_{\bf k}$  or $\omega_{\bf k}$, decoupled stable one-particle
dispersion relation of $\phi$. It is assumed that the real part of
$\tilde \omega^\pm ({\bf k})$ is positive. This consideration is similar
to the one considered in section \ref{stat} and thus, should yield the
same black-body radiation formula through the relation,
\begin{align}
\label{effectiveH}
\langle H \rangle_\beta \,=- \frac{\partial}{\partial \beta} \log
Z(\beta)\,, \qquad Z(\beta) = \int [d\phi] e^{ - \beta H}\,.
\end{align}

One may expand the field operator as
\begin{eqnarray} \label{thermalfield}
\phi({\bf x},\tau) =
\left(\frac{\beta}{V}\right)^{1/2}\sum_{n=-\infty}^\infty \int_{\bf p}
e^{i({\bf p \cdot x}+ \omega_n \tau)} \phi_n({\bf p})\,,
\end{eqnarray}
where $\beta \omega_n = 2 \pi n$ with $n$ integer so that $ \phi({\bf
x},\tau=0) = \phi({\bf x},\tau = \beta)$ and $\int_{\bf p}$ denotes the
momentum integral with an appropriate measure. Then the partition
function  is written in terms of the modes as,
\begin{align}
\label{logz}
\log Z &=- \mathrm {Tr} \log \beta
(\frac{\partial} {\partial \tau} + \tilde\omega^+({\bf p}))
(\frac{\partial} {\partial \tau} + \tilde\omega^-({\bf p}) ) \nn\\
   &=- \sum_{n=-\infty}^{\infty} \int_{\bf p} \log
   \Big( (2\pi n i +  \beta \tilde\omega_{\bf p}^+)
   (2\pi n  i + \beta \tilde\omega_{\bf p}^-)\Big)
\end{align}
where  $\beta \omega_n = 2 \pi n$  is used.
The thermal energy density is given as
\begin{align}
\langle H \rangle_\beta
 &= \sum_n \int_{\bf p}
   \Big( \frac {\tilde\omega_{\bf p}^+} { 2\pi n i +  \beta \tilde\omega_{\bf p}^+}
   +
  \frac {\tilde\omega_{\bf p}^-} { 2\pi n i+  \beta \tilde\omega_{\bf p}^-}  \Big) \nn
\\
  &=\int_{\bf p}
   \Big( \frac{\tilde\omega_{\bf p}^+ + \tilde\omega_{\bf p}^-}2
   + \frac {\tilde\omega_{\bf p}^+} { e^{\tilde\omega_{\bf p}^+} -1 }
   + \frac { \tilde\omega_{\bf p}^-} { e^{\tilde\omega_{\bf p}^-} -1}   \Big)
\end{align}
where the summation over the discrete $n$ is done
using the identities,
\begin{align}
\sum_{n=-\infty}^\infty \frac{1}{2\pi n i +  x}
&=  x\sum_{n=-\infty}^\infty \frac{1}{(2n\pi)^2 + x^2}
\nn
\\
\sum_{n=-\infty}^\infty \frac{1}{(2\pi n)^2+x ^2}
&=\frac{1}{|x|}\left(\frac12+\frac{1}{e^{|x|}-1}\right)\,.
\nn
\end{align}
Apart from the zero point energy (the first term in the thermal energy),
the thermal energy reproduces the the quantum mechanical result given
in section \ref{stat}.

In the above, we have constructed an effective thermal field theory in
the $\kappa-$Minkowski space-time, which reproduces the quantum
mechanical results given in Sec. II, by singling out the stable real
energy modes. This was possible since the quadratic part of the action
in momentum space is factorized into the product of each mode factors,
and thus the field operator is expanded as a linear combination of these
mode contributions. Following the same procedure one may include the
unstable modes also, as far as the dispersion relation of the decoupled
mode leads to the positive real energy, whose effective ``Hamiltonian"
is of the form,
\begin{align}
H[\phi] &= \int_x  \phi^\dagger(x) \prod_m \Big(\frac{\partial} {\partial
\tau} + \tilde \omega_m (-i\nabla) \Big) \phi(x)
\end{align}
where  $\tilde \omega_m ({\bf k})$ denotes the energy for the
diagonalized on-shell mode we are considering. One may reproduce this
effective ``Hamiltonian" from the expression of $\Delta_F ^{-1}(p) $ in
Eq.~(\ref{eq:deltaF-factorization}) or  $\Delta_S ^{-1}(p)  $ in
Eq.~(\ref{eq:DeltaS}) noting that
\begin{align}
\sinh x= x \prod_{n=1}^\infty \left(
1 + \frac {x^2}{n^2 \pi^2} \right) \,,\qquad
\cosh x= \prod_{n=0}^\infty \left(
1 + \frac {4 x^2}{(2n+1)^2 \pi^2} \right)\,.
\end{align}
Therefore, one may view Eq.~(\ref{effectiveH}) as the truncated version
of full spectra. By-product of this consideration is the solution to the
ambiguity of the integration  measure in the momentum integral: The
momentum integration in Eq.~(\ref{logz}) with the $\kappa$-deformed
Poincar\'e invariant measure is preferred even in this truncated
version. This is reasonable since the field theories in the
$\kappa-$Minkowski space-time and the related discussions of doubly
special
relativity~\cite{giovanni,girelli,freidel,amelino,livine,amelino2}, are
based on the symmetry of the system, namely, the $\kappa-$Poincar\'e
invariance. As we have discussed earlier, the field theory action and
related quantities are $\kappa-$Poincar\'e invariant in momentum space.
In this sense, the invariant measure in the momentum integral is needed.

%%%%%%%%%%%%%%%%%%%%%%%%%%%%%%%%%%%%%%%%%%%%%
\section{Summary and discussion}\label{sec:summary}
%%%%%%%%%%%%%%%%%%%%%%%%%%%%%%%%%%%%%%%%%%%%%%%
We have calculated the black body radiation spectra for massless scalar
particles in $\kappa-$Minkowski space-time, for particle systems of two
different dispersion relations corresponding to different orderings of
the kernel function of the Fourier transformation. We have shown that
the radiation spectrum depends on the form of the dispersion relations,
and thus on the ordering of the exponential kernel function, or on how
the $\kappa-$Poincar\'e algebra is realized in constructing field
theories in $\kappa-$Minkowski space-time.
 This implies that the
differently realized scalar field theories in $\kappa-$Minkowski
space-time are not physically equivalent.

The deviations in the radiation spectra from those of the commutative
case have two different origins. One is from the $\kappa-$dependent
modification of dispersion relations
and the other is due to the
presence of the nontrivial measure factor in the momentum integral,
all of which give the $O(T^6/\kappa^2)$
contribution to the thermal energy density.
As explained in the last section,
if we require the $\kappa-$deformed
Poincar\'e invariance in momentum space,
this measure contribution constitutes
the dominant correction to the radiation spectrum.

In Sec.III we have shown how different orderings of the exponential
kernel function lead to different realizations of the scalar field
theories in $\kappa-$Minkowski space-time. Noting that the scalar field
action in momentum space may be factorized into the product of each mode
factors, we have selected out the stable real energy modes to
second-quantize the scalar field theory for both the asymmetric and
symmetric ordering cases. We have used this fact in the last section, to
propose an effective thermal field theory, which reproduces the
black-body radiation spectra obtained by using the quantum mechanical
picture in Sec.II.

As noted in the last two sections, one of the most peculiar features of
the field theories in $\kappa-$Minkowski space-time, is
the  real and complex mode changing phenomenon.
That is, when momentum exceeds certain value,
$\kappa$ in the asymmetric ordering and 2$\kappa$ in the symmetric
ordering case, a real mode becomes complex and a complex mode becomes real.
To understand the role of  this new real mode,
we estimate  the contribution to the black-body radiation
spectra.

For asymmetric ordering case, the complex modes described by Eq.(51)
become real when $|{\bf k}|\geq \kappa$, satisfying the dispersion
relation, $ \tilde\Omega_{\bf k}= - \kappa \log(\frac{|{\bf
k}|}{\kappa}-1)$ for $\kappa <|{\bf k}|\leq 2\kappa$ and $
\tilde\Omega_{\bf k}= \kappa \log(\frac{|{\bf k}|}{\kappa}-1)$ for
$|{\bf k}|\geq 2\kappa$. These modes, in thermal equilibrium at
temperature $1/\beta$, will give internal thermal energy density,
\begin{align}
\rho_{a}^{1} (\beta)  &= \int_{2\kappa\geq|{\bf k}|\geq\kappa}
    \frac{d^3 {\bf k}\, e^{\alpha \tilde\Omega_{\bf
    k}/\kappa}}{(2\pi)^3}
    \frac{\tilde\Omega_{\bf k}^+}{e^{\beta \tilde\Omega_{\bf k}}-1}\,,
\qquad
\rho_{a}^{2} (\beta) = \int_{|{\bf k}|\geq 2\kappa}
    \frac{d^3 {\bf k}\, e^{\alpha \tilde\Omega_{\bf
    k}/\kappa}}{(2\pi)^3}
    \frac{\tilde\Omega_{\bf k}^+}{e^{\beta \tilde\Omega_{\bf
    k}}-1}\,.
\end{align}
Following the line of calculations in Sec. II, we find that the
contribution of these new high momentum modes to the energy density is
\begin{eqnarray} \label{anomal}
\rho_a =\rho_{a}^{1}+\rho_{a}^{2}= \frac{\kappa^2T^2}{\pi^2}\left[
    4\zeta(2)+\frac{\alpha
    \zeta(3)T}{\kappa}+\frac{6(2\alpha^2+9)\zeta(4)T^2}{\kappa^2}+\cdots
    \right] \,.
\end{eqnarray}
Note that the dominant contribution of this high momentum mode
is the term of $O(\kappa^2T^2)$, and does not
reduces to the commutative theory limit.

In the symmetric ordering case, the new high energy modes with $|{\bf
k}|\geq 2\kappa$ satisfies the dispersion relation
$ \omega_{a\bf k} = \kappa
    \log\left(\frac{{\bf k}^2}{2\kappa^2}-1+\frac{|{\bf
    k}|}{\kappa}\sqrt{\frac{\bf k^2}{4\kappa^2}-1}\right)$.
The contribution of these high momentum modes to the black-body
radiation spectrum turns out to be,
\begin{equation}\label{rho:sa}
\rho_{sa}^+ (\beta)+ \rho_{sa}^- (\beta)
    = \frac{2\kappa T^{3}}{\pi^2} \left[
    2\zeta(3)-\frac{6\zeta(4)T}{\kappa}
    + \frac{(12\bar \alpha^2+19)\zeta(5) T^2}{\kappa^2}
    +  O\left(\frac {T^3}{\kappa^3} \right) \right],
\end{equation}
whose dominant contribution is of the order
$O(\kappa T^3 )$ and does not have the correct commutative limit.

Field theories in $\kappa-$Minkowski space-time and the
$\kappa-$Poincar\'e algebra are constructed in such a way that the
formulations reduce to those in commutative space-time as $\kappa$
becomes infinite. Since the new high momentum modes do not satisfy this
criterion, one must impose a condition to suppress these modes from
appearing in the physical processes. One of the simplest ways to do this
is to restrict the momenta by $|{\bf k}|\leq \kappa$, which reminds us
of similar restriction needed in the formulations of doubly special
relativity~\cite{giovanni,livine}. In addition, the cut-off in the
momentum space affects the completeness of the energy spectra and the
effects on the coordinate space representation is to be seen.

Finally, in usual approach, the black body radiation is defined from the
radiation of a gauge field.
Therefore, to complete the study, we need to know how to construct the
$U(1)$ gauge field theory and then use the theory to calculate the black
body radiation.
Even though there exist some studies~\cite{dimit} on the $U(1)$ gauge
field theory in $\kappa-$Minkowski space-time, the study is still
incomplete and there are many things to be known.
We leave these subjects for later studies.

%\vspace{2cm}
\begin{appendix}

\section{bi-crossproduct basis of $\kappa-$Minkowski spacetime}

In this appendix we summarize the bi-crossproduct
basis of the $\kappa-$Poincar\'{e} algebra
and its  noncommutative Hopf algebra used in Sec. III
~\cite{majid}.
The ordinary Poincar\'e algebra is given as
\begin{align}
\label{ord-poincare}
\left[ M_{\mu\nu}, M_{\rho\sigma}\right]
&=i
\Big(\eta_{\mu\sigma}M_{\nu\rho}
+\eta_{\nu\rho}M_{\mu\sigma}
-(\mu \leftrightarrow \nu)
\Big) \nn \\
\left[ M_{\mu\nu}, p_{\rho}\right]
&=-i
\Big(\eta_{\mu\rho} p_{\nu}
- \eta_{\nu\rho} p_\mu \Big)
\nn \\
\left[p_{\mu}, p_{\nu}\right] &=0
\end{align}
where $\eta_{\mu\nu}=(+,-,-,-)$ is used.
In Eq.~(\ref{eq:Lorentz-rotation}),
 $M_i = \epsilon_{ijk}M_{jk}/2$
 and $N_i =M_{i0}$ are used.

The $\kappa-$deformed Poincar\'e algebra is
given by the modified form of the boost part
of Eq.~(\ref{ord-poincare}),
which depends on the ordering of exponential
operator $e^{-ip\cdot x}$,
Eq.~(\ref{NOrdering}). For the asymmetric ordering
 in section \ref{sec:asym-def},
the boost part is given as Eq.~(\ref{eq:boost});
\begin{align}
\label{asym-poincare}
\left[ M_{\mu\nu}, M_{\rho\sigma}\right]
&=i
\Big(\eta_{\mu\sigma}M_{\nu\rho}
+\eta_{\nu\rho}M_{\mu\sigma}
-(\mu \leftrightarrow \nu)
\Big)
\nn \\
\left[ M_i, p_j\right]
&=i \epsilon_{ijk} p_k \,,\quad
\left[ M_i, p_0\right] =0
\nn \\
\left[ N_i, p_j\right] & = i \delta_{ij} \Big( \frac
{\kappa}2 (1-e^{-2p^0/\kappa}) + \frac {{\bf p}^2} {2 \kappa} \Big)
-\frac{i}{\kappa} p_i p_j \,,\qquad \left[ N_i, p_0\right] =ip_i
\nn\\
\left[p_{\mu}, p_{\nu}\right] &=0
\end{align}
The corresponding
noncocommutative Hopf algebra (coalgebra, antipode, and counit) is
\begin{align}
\label{aym-coalgebra}
&\triangle(M_i) = M_i\otimes 1+ 1\otimes M_i,  \\
&\triangle(N_i) = N_i \otimes e^{-p_0/\kappa}+ 1\otimes
    N_i -\frac{1}{\kappa} \epsilon_{ijk}M_j\otimes  p_k, \nn \\
&\triangle(p_i) = p_i\otimes e^{-p_0/\kappa} + 1\otimes p_i, \nn \\
&\triangle(p_0) =  p_0\otimes 1 + 1\otimes p_0, \nn \\
&S(M_i) = -M_i, \nn\\
&S(N_i) = -\left(N_i+\frac{1}{\kappa}\epsilon_{ijk}
M_jp_k\right)e^{p_0/\kappa}, \nn \\
&S(p_i) = -p_i e^{p_0/\kappa}, \nn \\
&S(p_0) = -p_0, \nn \\
&\epsilon(p_\mu,M_i,N_i) =  0. \nn
\end{align}

The $\kappa-$Minkowski spacetime $T^*$ is defined as a dual Hopf algebra to the algebra of translations ($T$, momenta) with the use of
pairing,
\begin{eqnarray} \label{pairing}
\langle p_\mu, x_\nu\rangle = i \eta_{\mu\nu} ,
\end{eqnarray}
together with the axiom of Hopf algebra duality,
\begin{eqnarray} \label{duality}
\left.
\begin{tabular}{c}
$\langle t, x_\mu x_\nu\rangle = \langle t_{(1)}, x_\mu\rangle\,
    \langle t_{(2)}, x_\nu\rangle ,$ \\
$\langle p_\mu p_\nu, x\rangle = \langle p_\mu, x_{(1)}\rangle\,
    \langle p_\nu, x_{(2)}\rangle ,$ \\
\end{tabular}
\right\} ~~~ \forall p_\mu\in T, ~~x_\nu\in T^* .
\end{eqnarray}
From this we get
\begin{eqnarray} \label{comm x}
~[x_i,x_j] = 0, ~~~~[x_0, x_i] = \frac{i}{\kappa}x_i ,
\end{eqnarray}
and
\begin{eqnarray} \label{Delta x}
\triangle x_\mu = x_\mu \otimes  1+ 1\otimes x_\mu .
\end{eqnarray}
Using the Hopf algebra of spacetime coordinates
one writes down the
covariant action of $T$ on $T^*$ (module algebra):
\begin{eqnarray} \label{cov action}
t \triangleright x =  x_{(1)}\langle t,x_{(2)}\rangle , ~~~\forall x \in
T^*, t\in T \quad\quad t\triangleright (x y) = (t_{(1)} \triangleright
x) (t_{(2)}\triangleright y), \quad
    1 \triangleright x = x .
\end{eqnarray}

The action $\triangleleft $ of $h\in U(so(1,3))$ on momenta $p_\mu$
is defined by $p_\mu \triangleleft h= [h,p_\mu] $.
The duality relation of $\kappa-$Minkowski space given as
\begin{eqnarray} \label{h:x}
\langle t, h \triangleright x\rangle \equiv \langle t \triangleleft h,
x\rangle ,
   \quad  \forall t \in T, \quad h \in U(so(1,3)), \quad x\in T^* .
\end{eqnarray}
Then, Eq.~(\ref{asym-poincare})  is translated into
\begin{eqnarray} \label{MN on x}
M_i \triangleright x_0 &=&0, \quad M_i \triangleright x_j = i \epsilon_{ijk} x_k\,, \\
N_i \triangleright x_0 &=&i x_i, \quad N_i \triangleright x_j =i
\delta_{ij} x_0 \,.
\nn
\end{eqnarray}
From these, one has  the relations,
\begin{eqnarray} \label{M:x2}
N_i \triangleright x_0^2&=& (N_{i(1)}\triangleright x_0)
(N_{i(2)}\triangleright x_0) = i(x_0 x_i+x_i x_0) +
    \frac{x_i}{\kappa},  \\
N_i \triangleright x_j x_k&=& (N_{i(1)}\triangleright x_j)
(N_{i(2)}\triangleright x_k) = i(\delta_{ij} x_0 x_k+\delta_{ik} x_j
    x_0) + \frac{1}{\kappa}\left(\delta_{ij} x_k-\delta_{jk}x_i\right). \nn
\end{eqnarray}
Therefore, one has $N_i \triangleright (x_0^2-{\bf x}^2+3i x_0/\kappa) = 0$
and $x_0^2-{\bf x}^2+3i x_0/\kappa$ is a Lorentz-invariant.

Likewise, the $\kappa$-deformed Poincar\'e algebra is given for symmetric ordering given in section \ref{sec:sym-def}~\cite{amelino2}:
\begin{align}
\label{sym-poincare}
\left[ M_{\mu\nu}, M_{\rho\sigma}\right]
&=i
\Big(\eta_{\mu\sigma}M_{\nu\rho}
+\eta_{\nu\rho}M_{\mu\sigma}
-(\mu \leftrightarrow \nu)
\Big)
\nn \\
\left[ M_i, p_j\right]
&=i \epsilon_{ijk} p_k \,,\quad
\left[ M_i, p_0\right] =0
\nn \\
\left[ N_i, p_j\right] &= i \delta_{ij} \,
e^{\frac{p^0}{2\kappa}} \Big( \frac {\kappa}2 (1-e^{-2p^0/\kappa}) +
\frac {{\bf p}^2} {2 \kappa}  e^{-\frac{p^0}{\kappa}} \Big)
-\frac{i}{2\kappa} p_i p_j  e^{-\frac{p^0}{2\kappa}} \,,\qquad \left[
N_i, p_0\right] =ip_i e^{-\frac{p^0}{2\kappa}}
\nn\\
\left[p_{\mu}, p_{\nu}\right] &=0
\end{align}
and  its Hopf-algebraic structure,
\begin{align}
\label{sym-coalgebra}
&\triangle(M_i) = M_i\otimes 1+ 1\otimes M_i,  \\
&\triangle(N_i) = N_i \otimes e^{-p_0/\kappa}
+ 1\otimes N_i
 - \frac{1}{\kappa} \epsilon_{ijk}\, M_j
 	\otimes   p_k  e^{-p_0/(2\kappa)},\nn \\
&\triangle(p_i) =  p_i\otimes e^{-p_0/(2\kappa)}
    + e^{p_0/(2\kappa)}\otimes p_i\nn\\
&\triangle(p_0) =  p_0\otimes 1 + 1\otimes p_0, \nn \\
&S(M_i) = -M_i, \nn\\
&S(N_i) = -\left(N_i+\frac{1}{\kappa}\epsilon_{ijk}
M_j p_k  e^{-p_0/(2\kappa)} \right)e^{p_0/\kappa} \nn \\
&S(p_\mu) = -p_\mu  \nn\\
&\epsilon(p_\mu,M_i,N_i) =  0 \,. \nn
\end{align}

From this algebra one may find the duality relation
in coordinate space through the pairing~(\ref{pairing})
and  duality Eqs.~(\ref{duality},\ref{cov action},\ref{h:x}).
It turns out that the deformed algebra
in  the $\kappa$-Minkowski spacetime
does not change and has the same forms
Eqs.~(\ref{comm x},\ref{Delta x},\ref{MN on x},\ref{M:x2}) so that $x_0^2-{\bf x}^2+3i x_0/\kappa$ is Lorentz-invariant.

\end{appendix}

%%%%%%%%%%%%%%%%%%%%%%%%%%%%%%%%%%%%%%%%%%%%%%%%%%%%%%%%%%
\begin{acknowledgments}
This work was supported in part by Korea Science and
Engineering Foundation Grant (R01-2004-000-10526-0) and
by the Korea Research Foundation Grant funded by
Korea Government (MOEHRD, Basic Research Promotion Fund)
(KRF-2005-075-C00009; H.-C.K.) and  through the
the Center for Quantum Spacetime(CQUeST)
of Sogang University with grant (R11-2005-021).
\end{acknowledgments}
%%%%%%%%%%%%%%%%%%%%%%%%%%%%%%%%%%%%%%%%%%%%%%%%%%%%%%%%
%%%%%%%%%%%%%%%%%%%%%%%%%%%%%%%%%%%%%%%%%%%%%%%%%%%%%%%%%%


\begin{thebibliography}{10}

\bibitem{good}
C. L. Bennett {\it et al.}, Astrophys. J. Suppl. {\bf 148}, 1 (2003); D.
N. Spergel {\it et al.}, Astrophys. J. Suppl. {\bf 148}, 175 (2003); G.
Hinshaw {\it et al.}, Astrophys. J. Suppl. {\bf 148}, 135 (2003); S.
Sarkar, Nucl. Phys. B Proc. Suppl. {\bf 148}, 1 (2005).

\bibitem{doplicher}
S. Doplicher, K. Fredenhagen, and J. E. Roberts, Commun. Math. Phys.
{\bf 172}, 187 (1995).

\bibitem{wess}
P. Aschieri, C. Blohmann, M. Dimitrijevi\'{c},
F. Meyer, P. Schupp, and J. Wess, Class. Quant. Grav. {\bf 22}, 3511 (2005).

\bibitem{chaichian}
M. Chaichian, P. Kulish, K. Nishijima, A. Tureanu, Phys. Lett. B {\bf
604}, 98 (2004).



\bibitem{okon}
C. Chryssomalakos and E. Okon, Int. J. Mod. Phys. {\bf D 13,} 2003
[hep-th/0410211] (2004).

\bibitem{finetuning}
J.~Collins, A.~Perez, D.~Sudarsky, L~Urrutia, H~Vucetich, Phys.\ Rev.\
Lett.\ {\bf 93}, 191301 (2004); {\it and references there in}.

\bibitem{chamseddine}
A. H. Chamseddine, Int. J. Mod. Phys. {\bf A  16}, 759 (2001).


\bibitem{unitarity1}
D.~Bahns, S.~Doplicher, K.~Fredenhagen, G.~Piacitelli, Phys. Lett.\ {\bf
B 533}, 178 (2002).

\bibitem{unitarity2}
C.~Rim and J.~H.~Yee, Phys. Lett. {\bf B 574}, 111 (2003); J.\ Korean
Phys.\ Soc.\ {\bf 45}, 1435 (2004); C. Rim, Y. Seo, and J. H. Yee, Phys.
Rev. {\bf D 70}, 025006 (2004).

\bibitem{kappa}
J. Lukierski, A. Nowicki, H. Ruegg, and V. N. Tolstoy, Phys. Lett. {\bf
B 264}, 331 (1991); S. Majid and H. Ruegg, Phys. Lett. {\bf B 329}, 189
(1994).

\bibitem{kosinski}
P. Kosi\'{n}ski, J. Lukierski, and P. Ma\'{s}lanka,  Phys. Rev. {\bf D
62}, 025004 (2000).

\bibitem{glikman}
J. Kowalski-Glikman and S. Nowak, {\it Quantum $\kappa-$Poincar\'{e}
algebra from de Sitter Space of Momenta}, [hep-th/0411154].

\bibitem{daszkie}
M. Daszkiewicz, K. Imilkowska, J. Kowalski-Glikman and S. Nowak, Int. J.
Mod. Phys. {\bf A 20}, 4925 [hep-th/0410058] (2005).

\bibitem{majid}
S. Majid and H. Ruegg, Phys. Lett. {\bf B  334}, 348 (1994).

\bibitem{sitarz}
A. Sitarz, Phys. Lett. {\bf B 349}, 42 (1995).

\bibitem{gonera}
C. Gonera, P. Kosi\'{n}ski, and P. Ma\'{s}lanka, J. Math. Phys. {\bf
37}, 5820 (1996).

\bibitem{kim0}
H.-C. Kim, J. H. Yee, and C. Rim, Phys. Rev. {\bf D 75}, 045017 (2007).

\bibitem{freidel2}
L. Freidel, J. K.-Glikman, and S. Nowak, {\it From noncommutative
$\kappa-$Minkowski to Minkowski space-time}, [hep-th/0612170].

\bibitem{kim2}
H.-C. Kim, J. H. Yee, and C. Rim, Phys. Rev. {\bf D 72}, 103523 (2005).

\bibitem{fatollahi}
A. H. Fatollahi and M. Hajirahimi, Phys. Lett. {\bf  B 641}, 381 (2006);
Europhys. Lett. {\bf 75}, (2006).

\bibitem{amelino2}
A. Agostini, G Amelino-Camelia, and F. D'Andrea, Int. J. Mod. Phys. {\bf
A 19}, 5187 [hep-th/0306013] (2004); A. Agostini, F. Lizzi, A. Zampini,
Mod. Phys. Lett. {\bf A 17}, 2105 (2002) [hep-th/0209174]; A. Agostini,
hep-th/0312305 (Ph. D. Thesis); M. Daszkiewicz, J. Lukierski, and M.
Woronowicz, hep-th/0703200.

\bibitem{LRZ}
J.\ Lukierski, H.\ Ruegg and W.\ Zakrzewski, Ann.\ Phys.\ {\bf 243},
90-116 (1995).

\bibitem{realmodefieldtheory}
M. Arzano and A. Marciano, arXiv:0707.1329 [hep-th]; Phys. Rev. {\bf D 75}, 081701 [arXiv:hep-th/0701268].

\bibitem{giovanni}
Giovanni Amelino-Camelia, Phys. Lett. {\bf B 510}, 255 (2001)
[hep-th/0012238]; Int. J. Mod. Phys. {\bf D 11}, 35 (2002)
[gr-qc/0012051].

\bibitem{girelli}
F. Girelli and R. Livine, Braz. J. Phys. {\bf 35}, 432 [gr-qc/0412079]
(2005).


\bibitem{freidel}
L. Freidel, J. Kowalski-Glikman, L. Smolin, Phys. Rev. {\bf D 69},
044001 [hep-th/0307085] (2004).

\bibitem{amelino}
G Amelino-Camelia, {\it The three perspectives on the quantum-gravity
problem and their implications for the fate of Lorentz symmetry},
[gr-qc/0309054]; J. Kowalski-Glikman, {\it Intorduction to Doubly
Special Relativity}, [hep-th/0405273]; G Amelino-Camelia, L. Smolin, A.
Starodubtsev, Class. Quant. Grav. {\bf 13}, 3095 [hep-th/0306134]
(2004); Giovanni Amelino-Camelia, [gr-qc/0506117].

\bibitem{livine}
F. Girelli, E.R. Livine, and D. Oriti, Nucl. Phys. {\bf B 708}, 411
[gr-qc/0406100] (2005).

\bibitem{dimit}
M. Dimitrijevi\'{c}, L. Jonke, and L. M\"{o}ller, JHEP {\bf 0509}, 068
(2005) [hep-th/0504129]; P. N. Bikikov, J. Phys. A {\bf 31}, 6437 (1998)
[q-alg/9710019].


\end{thebibliography}
\end{document}